\documentclass[12pt]{iopart}

\usepackage{iopams}
\usepackage[english]{babel}
\usepackage{graphicx}

\begin{document}

%%%%%%%%%%%%%%%%%%%%%%%%%%%%%%%%%%%%%%%%%%%%%%%%%%%%%%%%%%%%%%%%%%%%%%%%
% HEADER
%%%%%%%%%%%%%%%%%%%%%%%%%%%%%%%%%%%%%%%%%%%%%%%%%%%%%%%%%%%%%%%%%%%
%
\title{Photon transport in binary photonic lattices}

\author{B.~M. Rodr\'{\i}guez-Lara, H. Moya-Cessa }
\address{INAOE, Coordinaci\'{o}n de \'{O}ptica, A.P. 51 y 216, 72000 Puebla, Puebla, M\'{e}xico.}
\ead{bmlara@inaoep.mx}

%%%%%%%%%%%%%%%%%%%%%%%%%%%%%%%%%%%%%%%%%%%%%%%%%%%%%%%%%%%%%%%%%%%%%%%%
% ABSTRACT
%%%%%%%%%%%%%%%%%%%%%%%%%%%%%%%%%%%%%%%%%%%%%%%%%%%%%%%%%%%%%%%%%%%
%
\begin{abstract}
We present a review on the mathematical methods used to theoretically study classical propagation and quantum transport in arrays of coupled photonic waveguides.
We focus on analysing two types of binary photonic lattices where self-energies or couplings are alternated.
For didactic reasons, we split the analysis in classical propagation and quantum transport but all methods can be implemented, \textit{mutatis mutandis}, in any given case.
On the classical side, we use coupled mode theory and present an operator approach to Floquet-Bloch theory in order to study the propagation of a classical electromagnetic field in two particular infinite binary lattices.
On the quantum side, we study the transport of photons in equivalent finite and infinite binary lattices by couple mode theory and linear algebra methods involving orthogonal polynomials.
Curiously the dynamics of finite size binary lattices can be expressed as roots and functions of Fibonacci polynomials.
\end{abstract}


\submitto{\PS}

%\maketitle

%%%%%%%%%%%%%%%%%%%%%%%%%%%%%%%%%%%%%%%%%%%%%%%%%%%%%%%%%%%%%%%%%%%%%%%%
% INTRODUCTION
%%%%%%%%%%%%%%%%%%%%%%%%%%%%%%%%%%%%%%%%%%%%%%%%%%%%%%%%%%%%%%%%%%%
%
\section{Introduction}
The analogy between linear lattices and the atom-field interaction
\cite{Crisp} or ion-laser interactions \cite{Jona,Tomb} has been a
fundamental step for the emulation, via classical interactions, of
quantum mechanical systems. This is important due to both pure
scientific interest and possible applications to quantum
computing. In the latter, the properties of classical systems have
been used to realize quantum computational operations by
quantum-like systems and, in particular, it has been show how a
controlled-NOT gate may be generated in nonhomogeneous optical
fibers \cite{Manko1}. At the fundamental level, \textit{e.g.}, it
has been possible to the emulate the most basic atom-field
interaction, the Jaynes-Cummings model,
theoretically~\cite{Longhi} and experimentally~\cite{Longhi2} with
arrays of photonic waveguides and, just to give another example,
it has been proposed to model non-linear coherent
states~\cite{Manko2} in waveguide arrays~\cite{Roberto}; linear
coherent states have also been modelled in linear arrays of
photonic waveguides~\cite{Keil2011p103601}.

In what follows we will take advantage of the simplest composite array of waveguides, \textit{i.e.}, binary arrays, to introduce the most basic mathematical methods used to study photonic lattices.
\Sref{sec:Model} gives an introduction to the two kinds of binary photonic lattices considered; those with either alternating self-energy or coupling.
Then, we proceed to study the propagation of classical electromagnetic fields through binary lattices of infinite size by use of coupled mode theory for lattices with alternating self-energy and by an operator approach to Floquet-Bloch theory in the case of an array with alternating couplings.
In \sref{sec:Quantum} we use coupled mode theory (with a twist given by the use of orthogonal polynomials) to solve a finite binary self-energies lattice in order to exemplify how the presented methods are valid, changing what needs to be changed, in all cases.
In this section we also find the dynamics of a finite binary couplings array by basic matrix theory methods.
Curiously, we find that the dynamics for both finite size binary lattices can be written in terms of Fibonacci polynomials evaluated at the roots of a Fibonacci polynomial which order is related to the size of the system.
In \sref{sec:Example} we introduce some quantities of interest when studying photon transport in arrays of photonic waveguides.
We show that initial states with a Gaussian distribution of amplitudes and linear coherent states, that is, initial states with Poisson-like distributions, reconstruct in binary lattices.
Finally, we present our conclusions.

%%%%%%%%%%%%%%%%%%%%%%%%%%%%%%%%%%%%%%%%%%%%%%%%%%%%%%%%%%%%%%%%%%%%%%%%
% MODEL
%%%%%%%%%%%%%%%%%%%%%%%%%%%%%%%%%%%%%%%%%%%%%%%%%%%%%%%%%%%%%%%%%%%
%
\section{Binary photonic lattices} \label{sec:Model}

\begin{figure}[htbp]
\centering
\includegraphics[scale=1]{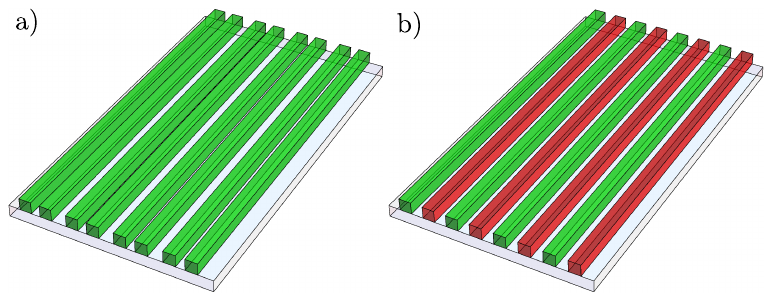}
\caption{(Color online) (a) Photonic binary super-lattice where identical waveguides alternate different nearest neighbor coupling. (b) Photonic binary super-lattice where homogeneously coupled waveguides alternate diffraction index.}
\label{fig:Fig1}
\end{figure}

A binary photonic super-lattice is composed by the repetition of a primitive unit cell where two different elements are characterized by one parameter.
One type of such binary waveguide arrays, shown in \fref{fig:Fig1}(a), is composed by identical waveguides where nearest neighbour coupling between them alternates between two values; hereby, we will refer to this type as binary coupling (BC) lattice.
In the other type, shown in \fref{fig:Fig1}(b), waveguides with two different refraction indexes alternate position and are homogeneously coupled; which we will call binary index (BI) lattice from here on.
The propagation of a classical light field in an infinite BC lattice is ruled by the scalar differential equation set for the field amplitude at the $j$th waveguide,
\begin{eqnarray} \label{eq:EqMotClassFieldB1}
\fl -\rmi \partial_{z} E_{j} = n E_{j} + g_{0} \left( E_{j+1} + E_{j-1} \right) +   (-1)^{j}  \delta \left( E_{j+1} - E_{j-1} \right), \quad j = -\infty, \ldots, \infty,
\end{eqnarray}
where the constant $n$ is the refraction index of each waveguide and a total coupling has been defined as $g_{0}= \left( g_{1} + g_{2} \right)$ with the difference as $\delta = \left(g_{1}  - g_{2}\right)/2$.
We have used the shorthand notation $\partial_{x}$ to express the partial derivative with respect to $x$.
An infinite BI lattice it is given by,
\begin{eqnarray} \label{eq:EqMotClassFieldB2}
\fl -\rmi \partial_{z} E_{j} = \left[ n_{0} + (-1)^{j}  \epsilon \right] E_{j} + g \left( E_{j+1} + E_{j-1} \right), \quad j = -\infty, \ldots, \infty,
\end{eqnarray}
where a base refraction index has been defined as $n_{0} = (n_{1} + n_{2})/2$ such that it is halfway between the refraction index at each waveguide, $n_{1}$ and $n_{2}$, \textit{i.e.}, $\epsilon = \vert n_{1} - n_{2} \vert /2$, and the tight binding coupling is given by the real constant $g$.
Infinite~\cite{Christodoulides1988p794,Sukhorukov2002p2112,Sukhorukov2005p1849,Vicencio2009p065801} and semi-infinite~\cite{Mihalache2007p10718} BC and BI lattices, with the addition of a non-linearity to each waveguide, are well known in the non-linear optics community where the existence and stability of diverse types of solitons have been studied along the years.

\begin{figure}[htbp]
  \centering
  \includegraphics[scale=1]{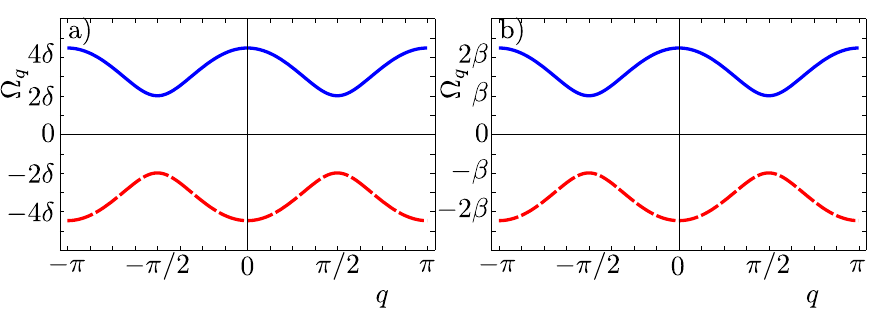}
  \caption{(Color online) Dispersion relation for infinite (a) BC with $g_{0}= 3 \delta$ and (b) BI lattices.}
  \label{fig:Fig2}
\end{figure}

In linear optics, it is possible to borrow from condensed matter theory a Floquet-Bloch result for quasiparticle motion on a chain in order to solve the evolution of a classical field, or a single photon, through these lattices~\cite{Kovanis1988p147}; \textit{e.g.}, the dispersion relation for a BC lattice, shown in \fref{fig:Fig2}(a),
\begin{eqnarray} \label{eq:FieldEquationsB1}
\Omega_{\phi}^2 =  4 \left[ \delta^2 + (g_{0}^2 - \delta^2) \cos^{2} \phi \right],
\end{eqnarray}
results of the infinite BC lattice has been used to discuss the existence of two propagation modes with opposite transverse velocities~\cite{Guasoni2008p1515}.
For a BI lattice the dispersion relation is,
\begin{eqnarray} \label{eq:FieldEquationsB2}
\Omega_{\phi}^2 =  \beta^2 + 4 \cos^{2} \phi, \quad \beta = \epsilon/g
\end{eqnarray}
The band gap structure of this dispersion relation, shown in \fref{fig:Fig2}(b), near the edge of the Brillouin zone,  $\phi_{B}= \pi / 2$, suggests the use of Bloch waves close to $\phi_{B}$ to simulate one-dimensional Dirac equations~\cite{Cannata1990p309}.
Under such a condition, a  photonic analogue of \textit{zitterbewegung} has been theoretically proposed~\cite{Longhi2010p235} and experimentally realized~\cite{Dreisow2010p143902}; more complex approaches of such a classical simulation include atoms in bichromatic optical lattices~\cite{Witthaut2011p033601,Szpak2011p050101(R)}.

%%%%%%%%%%%%%%%%%%%%%%%%%%%%%%%%%%%%%%%%%%%%%%%%%%%%%%%%%%%%%%%%%%%%%%%%
% PROPAGATION
%%%%%%%%%%%%%%%%%%%%%%%%%%%%%%%%%%%%%%%%%%%%%%%%%%%%%%%%%%%%%%%%%%%
%
\section{Classical electromagnetic field propagation} \label{sec:Classical}

\subsection{Coupled mode theory}

We are going to implement a coupled mode theory method on the infinite BC lattice described by \eref{eq:EqMotClassFieldB1}.
For starters, let us define the field amplitudes with a rotation proportional to the refractive index function, $n_{0}$, \textit{i.e.}, $E_{j} \rightarrow \rme^{\rmi n_{0} z} E_{j}$;
in this way, we can get rid of the term involving the refractive index function,
\begin{eqnarray} \label{eq:EqMotClassFieldB12}
\fl -\rmi \partial_{z} E_{j} =  g_{0} \left( E_{j+1} + E_{j-1} \right) +   (-1)^{j}  \delta \left( E_{j+1} - E_{j-1} \right), \quad j = -\infty, \ldots, \infty.
\end{eqnarray}
Now, we can define two auxiliary proper modes, say $A_{k} = \rme^{\rmi \Omega_{q} z} E_{2k}$ and $B_{k} = \rme^{\rmi  \Omega_{q} z} E_{2k-1}$, such that our differential set is now given by a coupled modes set
\begin{eqnarray}
(g_{0} + \delta) A_{k} + (g_{0} - \delta) A_{k+1} &=& \Omega_{q} B_{k}, \\
(g_{0} + \delta) B_{k} + (g_{0} - \delta) B_{k-1} &=& \Omega_{q} A_{k}.
\end{eqnarray}
It is trivial to find a three term recurrence relation for the mode $B$ from this coupled set of equations,
\begin{eqnarray}
\frac{A_{k-1} + A_{k +1}}{A_{k}} = \frac{\Omega_{\phi}^2 - 2 (g_{0}^{2} + \delta^{2})}{g_{0}^2 - \delta^2},
\end{eqnarray}
solved by
\begin{eqnarray}
A_{k} = \rme^{ \rmi 2 k \phi}, \\
B_{k} = \frac{2 \rme^{ \rmi (2 k+1) \phi}}{\Omega_{\phi}} \left[ g \cos \phi + \rmi \delta \sin \phi, \right]
\end{eqnarray}
under the restriction
\begin{eqnarray}
2 \cos 2 \phi = \frac{\Omega_{\phi}^2 - (g_{0}^{2} + \delta^{2})}{g_{0}^2 - \delta^2}.
\end{eqnarray}
From the expression above, we straightforwardly recover the dispersion relation
\begin{eqnarray}
\Omega_{\phi}^{2} = 4 \left[ \delta^2 + \left( g_{0}^2 - \delta^2 \right) \cos^2 \phi  \right].
\end{eqnarray}
Notice that we can also write the normal mode components and dispersion relation as,
\begin{eqnarray}
E_{j}^{(\phi)} = \rme^{\rmi j \phi}\left\{ \begin{array}{ll} 1 & j~\mathrm{even}, \\ \left(\frac{g_{0} \cos \phi + \rmi \delta \sin \phi}{ g \cos \phi - \rmi \delta \sin \phi }\right)^{\frac{1}{2}} & j~\mathrm{odd}, \end{array} \right. \\
\Omega_{\phi}^2 = \vert g \cos \phi + \rmi \delta \sin \phi\vert^2.
\end{eqnarray}
Under this treatment we can recover the field amplitude propagation at the $j$th waveguide for a field entering the lattice at the $m$th waveguide,
\begin{eqnarray}
E_{j}^{(m)} = \frac{1}{2 \pi} \int_{- \pi} ^{\pi} \rmd\phi ~\rme^{\rmi (m-j) \phi} \mathcal{E}_{j}^{(m)} , \\
\mathcal{E}_{j}^{(m)}= \rme^{-\imath \Omega_{\phi} z}  \times
\left\{
\begin{array}{ll}
1 & m-j ~\mathrm{even}, \\
\left(\frac{g_{0} \cos \phi - \rmi \delta \sin \phi}{ g \cos \phi + \rmi \delta \sin \phi }\right)^{\frac{1}{2}}  &j~\mathrm{even}, m~\mathrm{odd}, \\
\left(\frac{g_{0} \cos \phi + \rmi \delta \sin \phi}{ g \cos \phi - \rmi \delta \sin \phi }\right)^{\frac{1}{2}}  &j~\mathrm{odd}, m~\mathrm{even},
\end{array}
\right.
\end{eqnarray}
and we have found all the necessary information to study the propagation of any given initial classical field impinging a BC lattice.

\subsection{An operator approach to Floquet-Bloch theory}

It is possible to use an operator approach to find Floquet-Bloch waves.
In order to demonstrate it, let us write the differential equation set for an infinite BI lattice \eref{eq:EqMotClassFieldB2} in matrix form as
\begin{eqnarray} \label{eq:HamiltonianInfinite}
-\rmi \partial_{z} \bi{E} = \hat{H} \bi{E}, \\
\hat{H} = \left( -1 \right)^{\hat{n}} \beta  + \hat{V} + \hat{V}^{\dagger},
\end{eqnarray}
with solution
\begin{eqnarray}
\bi{E}(z) = \rme^{\rmi \hat{H} z } \bi{E}(0), \qquad \bi{E}(z) = \sum_{j=-\infty}^{\infty} E_{j}(t) \bi{E}_{j}.
\end{eqnarray}
The expression \eref{eq:HamiltonianInfinite} is identical to the whole differential set \eref{eq:EqMotClassFieldB2} up to a unitary transformation equivalent to a change into a frame rotating at a frequency proportional to the refractive index function $n_{0}$ and a change of units in terms of the homogeneous coupling $g$.
We used the unitary ladder operators, $\hat{V}^{\dagger}\hat{V} = \hat{V}\hat{V}^{\dagger} = 1$, defined as
\begin{eqnarray}
\hat{V} \vert k \rangle = \vert k-1 \rangle, \qquad
\hat{V}^{\dagger} \vert k \rangle = \vert k + 1 \rangle,\qquad
\hat{n} \vert k \rangle = k \vert k \rangle.
\end{eqnarray}
Where we substituted $\bi{E}_{k} \rightarrow \vert k \rangle$, \textit{i.e.}, the state $\vert k \rangle$ represents the field localized at the $k$th waveguide.
These ladder operators fulfil the commutation relations
\begin{eqnarray}
\left[ \hat{n},\hat{V} \right] =  -\hat{V}, \qquad
\left[ \hat{n},\hat{V}^{\dagger} \right] = \hat{V}^{\dagger}, \qquad
\left[ \hat{V}^{\dagger},\hat{V} \right] = 0.
\end{eqnarray}
By using this operator representation, we can define a phase state as
\begin{eqnarray}
\vert \phi \rangle = \sum_{k=-\infty}^{\infty} \rme^{\rmi k \phi} \vert k \rangle,
\end{eqnarray}
which is the Fourier transform of the localized field.
In other words, these operators allow us to do Floquet-Bloch theory,
\begin{eqnarray}
\vert j \rangle = \frac{1}{2 \pi} \int_{- \pi}^{\pi} \rmd \phi ~ \rme^{-\rmi j \phi} \vert \phi \rangle.
\end{eqnarray}
In this phase state basis, the action of the operator $\hat{H}_{BI}$ over some useful phase states reduces to:
\begin{eqnarray}
\hat{H} \vert \phi \rangle &=& \beta \vert \phi + \pi \rangle + 2 \cos \phi \vert \phi \rangle, \\
\hat{H}^2 \vert \phi \rangle &=& \left( \beta^2 + 4 \cos^{2} \phi \right) \vert \phi \rangle, \\
\hat{H}^2 \vert \phi + \pi \rangle &=&  \left( \beta^2 + 4 \cos^{2} \phi \right) \vert \phi + \pi \rangle,
\end{eqnarray}
From the last two equations it is possible to infer the dispersion relation,
\begin{eqnarray}
\Omega_{\phi}^{2} = \beta^2 + 4 \cos^{2} \phi.
\end{eqnarray}
In order to recover the rest of the information given by Floquet-Bloch theory, let us calculate the evolution of our phase state,
\begin{eqnarray}
\rme^{\rmi \hat{H} z} \vert \phi \rangle &=& \sum_{k = -\infty}^{\infty} \left[  \frac{(\rmi z)^{2 k}}{(2k)!} \left( \hat{H}^2 \right)^{k} + \frac{( \rmi z)^{2 k + 1}}{(2k+1)!} \left( \hat{H}^2 \right)^{k} \hat{H} \right] \vert \phi \rangle, \\
&=& \cos \Omega_{\phi} z \vert \phi \rangle + \rmi \left( \frac{\sin \Omega_{\phi} z}{\Omega_{\phi} }\right) \left( \beta \vert \phi + \pi \rangle + 2 \cos \phi \vert \phi \rangle \right).
\end{eqnarray}
Now, if the field started at the $m$th waveguide, $\vert \Psi(0) \rangle = \vert m \rangle $, one can write the field amplitude at the $j$th waveguide as $E_{j}^{(m)}= \langle j \vert m(t) \rangle$ and, by use of $\langle m \vert \phi \rangle = \rme^{\rmi m \phi}$, recover the field amplitude evolution from standard Floquet-Bloch theory
\begin{eqnarray}
E_{j}^{(m)} &=& \frac{1}{2 \pi} \int_{-\pi}^{\pi} \rmd\phi ~ \rme^{\rmi \left( j-m \right) \phi} \mathcal{E}_{j}^{(m)}, \\
\mathcal{E}_{j}^{(m)} &=&  \cos \Omega_{\phi} z  + \rmi \left( \frac{ 2  \cos \phi + \beta \rme^{\rmi (m-j) \pi} }{\Omega_{\phi} }\right) \sin \Omega_{\phi} z  .
\end{eqnarray}

%%%%%%%%%%%%%%%%%%%%%%%%%%%%%%%%%%%%%%%%%%%%%%%%%%%%%%%%%%%%%%%%%%%%%%%%
% TRANSPORT
%%%%%%%%%%%%%%%%%%%%%%%%%%%%%%%%%%%%%%%%%%%%%%%%%%%%%%%%%%%%%%%%%%%
%
\section{Photon transport} \label{sec:Quantum}

In some cases one has to consider the evolution of a quantum field in an array of coupled photonic waveguides; in a general case, it could be arrays of coupled microring resonators, coupled cavities, or capacitively coupled strip-line resonators.
When such a system is presented it is usually described by a Hamiltonian which, in the case of finite BC and BI lattices of size $N$, is given by
\begin{eqnarray}
\hat{H}_{BC} =  \sum_{j=0}^{N-1} \left[ g_{0} - (-1)^{j} \delta \right] \left(  \hat{a}_{j} \hat{a}_{j+1}^{\dagger} + \hat{a}_{j}^{\dagger} \hat{a}_{j+1} \right),\\
\hat{H}_{BI} = \sum_{j=0}^{N} \left( -1 \right)^{j} \beta   ~\hat{a}_{j}^{\dagger} \hat{a}_{j} + \sum_{j=0}^{N-1} \left(  \hat{a}_{j} \hat{a}_{j+1}^{\dagger} + \hat{a}_{j}^{\dagger} \hat{a}_{j+1} \right),
\end{eqnarray}
where we have moved into an adequate reference frame rotating at a frequency defined by the refractive index function $n_{0}$ and set units in terms of $\hbar$ and $\hbar g$, respectively. The operator $\hat{a}^{\dagger}_{k}$ ($\hat{a}_{k}$) creates (annihilates) a photon in the $k$th waveguide.

In the Heisenberg picture, the equations of motion for these systems are
\begin{eqnarray} \label{eq:EqMotAnnihilation}
 \textrm{BC}: \qquad \rmi \partial_{t} a_{j} &=&  g_{0} \left( \hat{a}_{j+1} +\hat{a}_{j-1} \right) + \left( -1 \right)^{j}  \left( \hat{a}_{j+1} -\hat{a}_{j-1} \right), \\
 \textrm{BI}: \qquad \rmi \partial_{t} a_{j} &=&  \left( -1 \right)^{j} \beta \hat{a}_{j} + \hat{a}_{j-1} + \hat{a}_{j+1},
\end{eqnarray}
these sets with $j=0,1,\ldots, N,$ are the finite equivalent, by making the substitutions $\hat{a}_{j} \rightarrow - E_{j}$ alongside $t \rightarrow z$ , of the differential sets ruling the propagation of a classical field through the corresponding binary photonic lattice in  \eref{eq:EqMotClassFieldB12} and \eref{eq:HamiltonianInfinite}, respectively.

\subsection{Coupled mode theory}

Let us start with the finite BC lattice and introduce the transformation
\begin{eqnarray}
\hat{T} = \rme^{\rmi g_{0} \sum_{j=0}^{N-1} \left(  \hat{a}_{j} \hat{a}_{j+1}^{\dagger} + \hat{a}_{j}^{\dagger} \hat{a}_{j+1} \right)},
\end{eqnarray}
such that a general states is defined by $\vert \psi \rangle = \hat{T} \vert \phi \rangle$ and leads to the Schr\"rodinger equation in units of $\hbar \delta$
\begin{eqnarray}
\rmi \partial_{t} \vert \phi \rangle = \tilde{H} \vert \phi \rangle,\\
\tilde{H} = \sum_{j=0}^{N-1} - (-1)^{j} \left(  \hat{a}_{j} \hat{a}_{j+1}^{\dagger} + \hat{a}_{j}^{\dagger} \hat{a}_{j+1} \right),
\end{eqnarray}
with equations of motion in the Heisenberg picture
\begin{eqnarray}
\rmi \partial_{t} a_{j} &=&   \left( -1 \right)^{j}  \left( \hat{a}_{j+1} -\hat{a}_{j-1} \right), \qquad j=0, 1, \ldots, N.
\end{eqnarray}
Now, let us define the coupled modes as $\hat{a}_{2k}(t) = - \rmi \rme^{\Omega t} \hat{b}_{2k+1}$ and $\hat{a}_{2k+1}(t) =  \rme^{\Omega t} \hat{b}_{2k+2}$.
Notice that we have used $\Omega t$ in the argument of the exponential instead of $\rmi \Omega t$, thus we are looking for a purely imaginary $\Omega$.
This allows us to write an eigenequation in the form,
\begin{eqnarray}
M \bi{b} = \Omega \bi{b}, \qquad M_{i,j} = \delta_{i,j-1} + \delta_{i-1,j}, \qquad \bi{b} = \sum_{j=0}^{N-1} c_{j} \hat{\bi{b}}_{j+1},
\end{eqnarray}
where it is possible to recover a recurrence relation for the coefficients of the normal modes,
\begin{eqnarray}
c_{2} = \Omega c_{1}, \\
c_{3} = \Omega c_{2} + c_{1}, \\
\vdots \\
\Omega c_{N} + \Omega c_{N-1}  = 0,
\end{eqnarray}
that is solved by Fibonacci polynomials,
\begin{eqnarray}
c_{j} = F_{j}(\Omega).
\end{eqnarray}
The last of the recurrence relations gives a boundary condition that translates into the expression $F_{N+1}(\Omega) = 0$ that is solved by~\cite{Hoggatt1973p271}
\begin{eqnarray}
\Omega(k) = \left\{ \begin{array}{ll}
\pm 2 \rmi \sin \frac{(2k + 1) \pi}{N+1} & \mathrm{even~} N, \\
\pm 2 \rmi \sin \frac{k \pi}{N+1} & \mathrm{odd~} N,
\end{array}
\right.
\end{eqnarray}
and we recover the purely imaginary eigenvalue that we were looking for.
So, the normal modes $\hat{c}_{k} = \sum_{j=0}^{(N+1)/2} \rmi F_{2j+1}(\rmi \lambda_{k}) \hat{a}_{2j} + F_{2j+2}(\rmi \lambda_{k}) \hat{a}_{2j+1}$ (up to a normalization constant) with eigenvalues $\lambda_{k} = -\rmi \Omega(k)$, diagonalize the Hamiltonian $\tilde{H} = \sum_{j=0}^{N-1} \lambda_{k} \hat{c}_{k} \hat{c}_{k}^{\dagger}$ and we have found the  dynamics of the system.
It is trivial to go back into the original frame.

\subsection{A linear algebra approach}

Now, let us consider a finite BC lattice.
Heisenberg equations of motion for this system may be written as the matrix differential equation,
\begin{eqnarray}
\fl \partial_{t} \bi{a} = - \rmi M \bi{a},  \qquad M_{ij} = (-1)^{i} \beta \delta_{i,j} + \delta_{i, j-1} + \delta_{i, j+1}, \qquad \bi{a}(t) = \rme^{- \rmi M t} \bi{a}(0),
\end{eqnarray}
The solution to this system may be found in a number of ways
\cite{Dodo} and we do it  by finding the  system, $\{V,\Lambda\}$,
of the matrix $M = V \Lambda V^{\dagger}$
\cite{RodriguezLara2011p053845}; where the eigenvalues matrix
$\Lambda$ is a matrix which diagonal elements are the $N$
eigenvalues of the matrix $M$, $\{ \lambda_{j} \}$, and the
eigenvector matrix $V$ is a matrix which rows, $\vec{v}_{j}$, are
the $N$ eigenvectors of M.

The characteristic polynomial, $p_{N}$, of tridiagonal matrix $M$ is found via the method of minors~\cite{Horn1990} as
\begin{eqnarray}
p_{N}(\lambda) &=& \left\{ \begin{array}{ll}
(-1)^{N/2} F_{N+1}(\sqrt{\beta^{2} - \lambda^{2}}) & N ~\mathrm{even},  \\
(-1)^{(N-1)/2} \frac{\beta - \lambda}{\sqrt{\beta^{2} - \lambda^{2}}}  F_{N+1}(\sqrt{\beta^{2} - \lambda^{2}}) & N ~\mathrm{odd},
\end{array} \right. \\
&=& \left\{ \begin{array}{ll}
(-1)^{N/2} ~b_{N/2}(\beta^{2} - \lambda^{2}) & N ~\mathrm{even},  \\
(-1)^{(N-1)/2} \left(\beta - \lambda\right) ~B_{(N-1)/2}(\beta^{2} - \lambda^{2}) & N~\mathrm{odd},
\end{array} \right.
\end{eqnarray}
where $F_{n}(x)$ is the $n$th Fibonacci polynomial \cite{Hazewinkel2002}, $b_{n}(x)$ and $B_{n}(x)$ are $n$th Morgan-Voyce polynomials~\cite{MorganVoyce1959p321}.
The roots of $F_{N+1}(x)=0$ are well known \cite{Hoggatt1973p271} and yield the eigenvalues,
\begin{eqnarray} \label{eq:EigenvaluesFinite}
\lambda_{j} = \left\{ \begin{array}{ll}
 -\sqrt{\beta^2 + 4 \cos^{2} \left( \frac{j \pi}{N+1}\right)} & j \le N/2,\\
 \sqrt{\beta^2 + 4 \cos^{2} \left( \frac{j \pi}{N+1}\right)} &  j > N/2.\\
 \end{array} \right.
\end{eqnarray}
These proper values, alongside $\left( M - \lambda_{j} \mathbb{I} \right)\vec{v}_{j} =0$, deliver recurrence relations fulfilled by eigenvector components,
\begin{eqnarray}
V_{j,k} &=& \frac{u_{j,k}}{\sqrt{\sum_{k=0}^{N-1} u_{j,k}^{2}}}, \quad j,k=0,\ldots,N-1, \label{eq:EigenfunctionsFinite}
\end{eqnarray}
where orthogonal polynomials, $u_{j,k}$, are defined as:
\begin{eqnarray} \label{eq:EigenfunctionsFiniteEven}
u_{j,k} &=& \left\{ \begin{array}{ll}
(-1)^{k/2} F_{k+1}\left( 2 \rmi \left\vert \cos \frac{j \pi}{N+1} \right\vert\right) & k ~\mathrm{even},  \\
- \rmi (-1)^{(k+1)/2} \frac{\beta - \lambda_{j}}{2 \left\vert \cos \frac{j \pi}{N+1} \right\vert}  F_{k+1}\left( 2 \rmi \left\vert \cos \frac{j \pi}{N+1} \right\vert\right) & k ~\mathrm{odd} ,
\end{array} \right. \\
&=& \left\{ \begin{array}{ll}
(-1)^{k/2} ~b_{k/2}\left( -4 \cos^{2} \frac{j \pi}{N+1} \right) & k ~\mathrm{even},  \\
(-1)^{(k+1)/2} \left(\beta - \lambda_{j}\right) ~B_{(k-1)/2}\left( -4 \cos^{2} \frac{j \pi}{N+1}  \right) & k ~\mathrm{odd} .
\end{array} \right.
\end{eqnarray}
Thus the Hamiltonian is diagonalized, $\hat{H} = \sum_{j=0}^{N-1} \lambda_{j} \hat{c}_{j}^{\dagger} \hat{c}_{j}$, in terms of delocalized normal modes $\bi{c} = V^{\dagger} \bi{a}$, giving a time evolution $ \bi{a}(t) = \rme^{-\rmi M t} \bi{a}(0) = V \rme^{-\rmi \Lambda t} V^{\dagger} \bi{a}(0)$.

%%%%%%%%%%%%%%%%%%%%%%%%%%%%%%%%%%%%%%%%%%%%%%%%%%%%%%%%%%%%%%%%%%%%%%%%
% QUANTITIES
%%%%%%%%%%%%%%%%%%%%%%%%%%%%%%%%%%%%%%%%%%%%%%%%%%%%%%%%%%%%%%%%%%%
%
\section{Quantities of interest} \label{sec:Example}

For the sake of space and simplicity we will only discuss typical quantities of interest when studying the propagation of a quantum electromagnetic field in an array of photonic waveguide lattices.
It is trivial to show that the time evolution of an arbitrary initial state can be obtained from the evolution of the annihilation operator found in the last section.
In order to generalize, let us write such an evolution as
\begin{eqnarray}
\hat{a}_{j} = \sum_{k=0}^{N-1} U_{j,k}(t) \hat{a}_{k}(0),
\end{eqnarray}
where it is trivial to construct the matrix $U(t)$ from the methods given in the last section.

The most basic and visually appealing quantity of interest is the transport of $m$ photons impinging a single-waveguide at time zero,
\begin{eqnarray}
\vert \psi_{p}(0) \rangle = \frac{1}{\sqrt{m!}} \hat{a}^{\dagger m}_{p}(0) \vert 0 \rangle, \qquad p =0, \ldots, N-1.
\end{eqnarray}
The mean photon number at the $q$th waveguide for such an initial state after propagation is given by
\begin{eqnarray}
\langle \hat{n}_{q} \rangle_{p} = \langle \psi_{p}(0) \vert \hat{a}(t)_{q}^{\dagger} \hat{a}(t)_{q} \vert \psi_{p}(0) \rangle = m \vert U_{p,q}(t) \vert^2.
\end{eqnarray}
In the case of single photon input, $m=1$, the expression for the mean photon number is identical to the normalized intensity at the $q$th waveguide for a propagating classical electromagnetic  field.

\begin{figure}[htbp]
\centering
\includegraphics[scale=1]{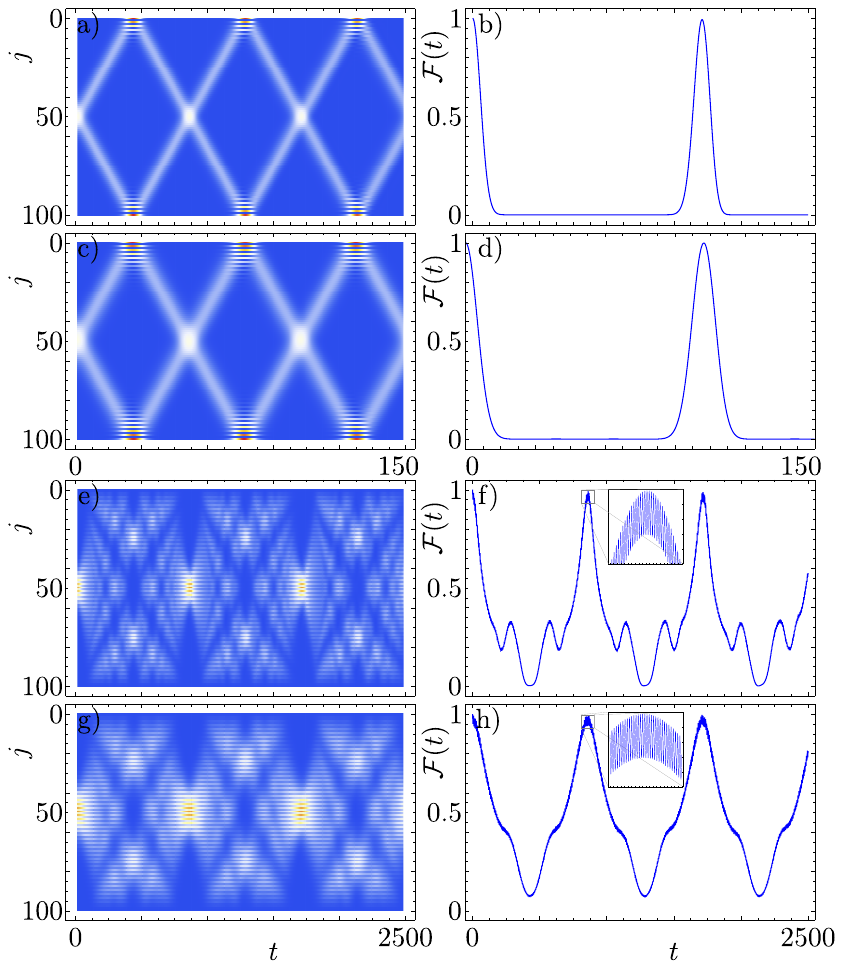}
\caption{(Color online). Time evolution of (a,c,e,f) the mean photon number for (a,e) Gaussian- and (c,g) Poisson-like input distributions with parameters $\left\{\omega_{0}=7,~q=0.55 \pi \right\}$ and $\alpha = \sqrt{50}$, in that order, and (b,d,f,h) their respective fidelities in a (a-d) BC and (e-h) BI with $\beta=0.5$ lattice of size $N=101$.  Time in units of $\delta$ for BC and $g$ for BI lattice.}
\label{fig:Fig3}
\end{figure}

Superposition states of single photons are also interesting,
\begin{eqnarray}
\vert \psi_{s}(0) \rangle = \sum_{j=0}^{N-1} \alpha_{j} \hat{a}^{\dagger}_{j}(0) \vert 0 \rangle, \qquad \sum_{j=0}^{N-1} \vert \alpha_{j} \vert^2 = 1,
\end{eqnarray}
and we can easily calculate the probabilities of finding the photon in the $q$th waveguide,
\begin{eqnarray}
\langle \hat{n}_{q} \rangle_{s} =  \left\vert \sum_{j=0}^{N-1} \alpha_{j} U_{p,q}(t) \right\vert^2.
\end{eqnarray}
Here, we are interested in two peculiar single-photon distributions.
One that we will call a \textit{Gaussian-like} input distribution,
\begin{eqnarray}
\vert \psi(0) \rangle = \sum_{j=0}^{N-1} e^{- k_{j}^{2} /(2 w_{0}^2) + \imath q k_{j}/2} \hat{a}_{j}^{\dagger} \vert 0 \rangle, \quad k_{j} = j-\lceil N/2 \rceil,
\end{eqnarray}
which could be thought as the result of a Gaussian beam whose intensity peak is aligned with the center of the lattice impinging parallel, $q=0$, or at an angle, $q\ne0$, with the propagation axis \cite{Longhi2010p235,Dreisow2010p143902}.
The other is similar to a quantum coherent state and is produced by a single-photon entering a Glauber-Fock lattice at the zeroth waveguide \cite{PerezLeija2010p2409,RodriguezLara2011p053845}, we will call this a \textit{Poisson-like} input distribution,
\begin{eqnarray}
\vert \alpha \rangle = e^{- \vert \alpha \vert^{2}/2} \sum_{j=0}^{N-1} \frac{\alpha^{j}}{\sqrt{j!}} \hat{a}_{j}^{\dagger} \vert 0 \rangle.
\end{eqnarray}
In these Gaussian- and Poisson-like distributions, the maximum probability to find the single-photon is given at the $\lceil N/2 \rceil$th and $\lceil\vert\alpha\vert^2\rceil$th waveguides (where $\lceil x \rceil$ has been used to express $x$ rounded up), in that order, and the variance is given by $w_{0}$ and $\lceil N/2 \rceil$, respectively.
Both initial states have a \textit{momentum} in the direction perpendicular to propagation given by $q$ and $\mathrm{Im}(\alpha)$ in the sense that the center of mass for an input with $q\ne0$ or $\mathrm{Im}(\alpha) \ne 0$ will drift to the right or left side of the lattice depending on the sign of $q$ or $\mathrm{Im}(\alpha)$.
Notice that a Gaussian-like distribution will always be symmetric with respect to the center of the lattice, which is not the case for the coherent-like distribution.
Propagation of Gaussian-like input distributions have been studied in infinite BC and BI lattices. In the former, the lattice splits the $q=0$ input in two components propagating in opposite directions \cite{Guasoni2008p1515}. In the latter, BI lattices allow the clasical simulation of the Dirac equation when the parameter q is close to the edge of the Brillouin zone \cite{Longhi2010p235,Dreisow2010p143902}.
To our knowledge there exist no record in the literature of propagation of Poisson-like input distributions through finite binary lattices.
\Fref{fig:Fig3} shows the mean photon probability for two kinds of superposition states impinging a BI and BC lattice, respectively.
It is possible to see that the states reconstruct periodically in both cases through the fidelity function
\begin{eqnarray}
\mathcal{F}(t) = \vert \langle \psi_{s}(0) \vert \psi_{s}(t) \rangle \vert^2.
\end{eqnarray}

\begin{figure}[htbp]
  \centering
  \includegraphics[scale=1]{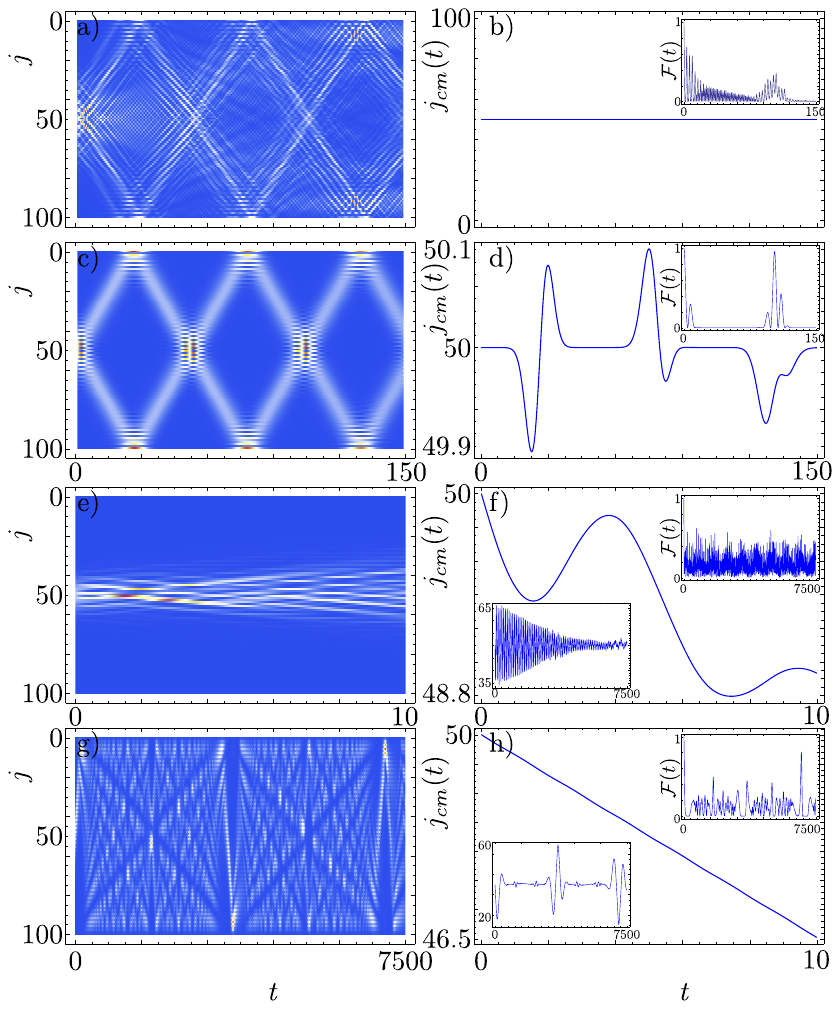}
  \caption{(Color online). Time evolution of (a,c,e,f) the mean photon number for (a,e) Gaussian- and (c,g) Poisson-like input distributions with complex parameters $\left\{ \omega_{0}=7,~q=0.55 \pi\right\}$ and $\alpha = \sqrt{50-0.55\pi} + \imath \sqrt{0.55 \pi}$, in that order, and (b,d,f,h) their respective center of mass (with top-right insets showing the fidelities and top-bottom a longer-time evolution of the center of mass) in a (a-d) BC and (e-h) BI with $\beta=0.5$ lattice of size $N=101$.  Time in units of $\delta$ for BC and $g$ for BI lattice.}
  \label{fig:Fig4}
\end{figure}

It is also possible to study the time evolution of the center of mass,
\begin{eqnarray} \label{eq:CenterMass}
j_{cm} = \sum_{k=0}^{N-1} k \langle \hat{a}_{k}^{\dagger} \hat{a}_{k} \rangle,
\end{eqnarray}
for Gausian- and Poisson-like input distributions with complex parameters.
Time evolution of the Fidelity for Gaussian-like distributions heavily fluctuates with values below $1/2$, see top-right insets in \fref{fig:Fig4}(b,f), and its center of mass localizes at the central waveguide after a long time; after $t=7500 g$ in the case shown in \fref{fig:Fig4}(f).
Only Poisson-like distributions partially reconstruct, see top-right insets in \fref{fig:Fig4}(d,h).
Of course, well-known results regarding classical simulation of Dirac equation \cite{Longhi2010p235} are reproduced; for instance optical \textit{zitterbewegun} by using a Gaussian-like distribution, \fref{fig:Fig4}(f).
Notice that the center of mass of the coherent-like distribution, \fref{fig:Fig4}(h), wobbles with an almost negligible amplitude compared to that of the Gaussian-like distribution, \fref{fig:Fig4}(f), and it appears to reconstruct with low fidelity after the second edge reflection; top-right inset in \fref{fig:Fig4}(h).

Another interesting set of states are product states,
\begin{eqnarray}
\vert \psi_{ps}(0) \rangle = \prod_{j=1}^{k} \hat{a}_{x_{k}}^{\dagger} \vert 0 \rangle,
\end{eqnarray}
where $\bi{x} = (x_{1},\ldots,x_{k})$ with $x_{i} \in [0,N-1]$ and $x_{i} \neq x_{j}$ for any $i \neq j$.
The evolution of the photon number at the $q$th waveguide for product states is given by
\begin{eqnarray}
\langle \hat{n}_{q} \rangle_{ps} = \sum_{j=1}^{k} \vert U_{x_{j},q} \vert^{2},
\end{eqnarray}
An example of these states is the two-photon product state given by
\begin{eqnarray}
\vert \psi_{ps}(0) \rangle =  \hat{a}_{j}^{\dagger} \hat{a}_{k}^{\dagger} \vert 0 \rangle,
\end{eqnarray}
that gives a mean photon number evolution and two-photon correlation function
\begin{eqnarray}
\langle \hat{n}_{q} \rangle_{ps} =  \vert U_{j,q} \vert^{2} + \vert U_{k,q} \vert^{2}, \\
\Gamma_{p,q}^{(ps)} = \vert U_{p,j} U_{q,k} + U_{p,k} U_{q,j} \vert^2 .
\end{eqnarray}
For the sake of space we will finish this section presenting how to deal with NOON states.
A higher order NOON state, their mean-photon evolution at the $q$th waveguide, and two-photon correlation are given by the expression,
\begin{eqnarray}
\vert \psi(0) \rangle = \frac{1}{2 \sqrt{m!}} \left( \hat{a}_{j}^{\dagger m} + \rme^{\rmi m \phi }\hat{a}_{k}^{\dagger m}  \right) \vert 0 \rangle, \qquad m= 2, 3, \ldots, \\
\langle n_{q} \rangle = \frac{m}{2} \left( \vert U_{j,q}  \vert^2  + \vert U_{k,q} \vert^2 \right), \\
\Gamma_{p,q} = \vert U_{p,j} U_{q,j} \vert^2 + \vert U_{p,k} U_{q,k} \vert^2 + 2\mathrm{Re}\left(\rme^{\rmi m \phi}  U_{p,j}^{\ast} U_{q,j}^{\ast} U_{p,k} U_{q,k} \right).
\end{eqnarray}

%%%%%%%%%%%%%%%%%%%%%%%%%%%%%%%%%%%%%%%%%%%%%%%%%%%%%%%%%%%%%%%%%%%%%%%%
% CONCLUSION
%%%%%%%%%%%%%%%%%%%%%%%%%%%%%%%%%%%%%%%%%%%%%%%%%%%%%%%%%%%%%%%%%%%
%
\section{Conclusion} \label{sec:Conclusion}

We presented a review on mathematical methods used to study infinite and finite photonic lattices and the quantities of interest for quantized electromagnetic fields.
Each method is presented in a particular context but any of the methods can be used in all case by making the necessary alterations.
Alongside this review, we introduced a novel result, up to our knowledge, in the form of the exact spectra and proper modes for BC and BI lattices of size $N$ (BI latices include the single-type lattice when the characteristic parameter is zero) in terms of the roots of the $N+1$ Fibonacci polynomial and the Fibonacci polynomials evaluated at these roots, in that order.
In order to illustrate our results we chose to focus on the evolution of multiple-waveguide single-photon inputs in the form of what we called Gaussian- and Poisson-like distributions impinging odd lattices; these distributions are related to Gaussian beams and to the output from Glauber-Fock lattices, respectively.
Due to their spectral decompositions, Gaussian- and Poisson-like distributions with real parameters partially reconstruct quasi-periodically when they impinge a binary super-lattice with their intensity peak aligned with the middle of the lattice.

\ack HMC acknowledges L. Moya-Rosales involvement in the first stage of this work.
BMRL is grateful to Miguel Bandres for useful discussion and comments.

%%%%%%%%%%%%%%%%%%%%%%%%%%%%%%%%%%%%%%%%%%%%%%%%%%%%%%%%%%%%%%%%%%%%%%%%
% REFERENCES
%%%%%%%%%%%%%%%%%%%%%%%%%%%%%%%%%%%%%%%%%%%%%%%%%%%%%%%%%%%%%%%%%%%
%
\section*{References}
\bibliographystyle{unsrt}
%\bibliography{E:/Bibliography/references}

\end{document}